\documentclass[sigconf]{acmart}

\AtBeginDocument{%
  \providecommand\BibTeX{{%
    \normalfont B\kern-0.5em{\scshape i\kern-0.25em b}\kern-0.8em\TeX}}}

\usepackage{xcolor}

\begin{document}
\title{With Great Power Comes Great Responsibility: The Role of Software Engineers}

\author{Stefanie Betz}
\orcid{0000-0002-3613-5893}
\affiliation{%
  \institution{Furtwangen University and LUT University}
  \streetaddress{Robert-Gerwig-PLatz 1}
  \city{Furtwangen}
  \country{Germany and Finland}}
\email{besi@hs-furtwangen.de}

\author{Birgit Penzenstadler}
\orcid{0000-0002-5771-0455}
\affiliation{%
  \institution{Chalmers University of Technology and LUT University}
  \streetaddress{Chalmersplatsen 4}
  \city{Gothenburg}
  \country{Sweden and Finland}}
\email{birgitp@chalmers.se}

\renewcommand{\shortauthors}{Betz and Penzenstadler, et al.}

\begin{abstract}
The landscape of software engineering is evolving rapidly amidst the digital transformation and the ascendancy of AI, leading to profound shifts in the role and responsibilities of software engineers. This evolution encompasses both immediate changes, such as the adoption of Language Model-based approaches in coding, and deeper shifts driven by the profound societal and environmental impacts of technology. Despite the urgency, there persists a lag in adapting to these evolving roles. By fostering ongoing discourse and reflection on Software Engineers role and responsibilities, this vision paper seeks to cultivate a new generation of software engineers equipped to navigate the complexities and ethical considerations inherent in their evolving profession.
\end{abstract}



\begin{teaserfigure}\centering
\includegraphics[width=12cm]{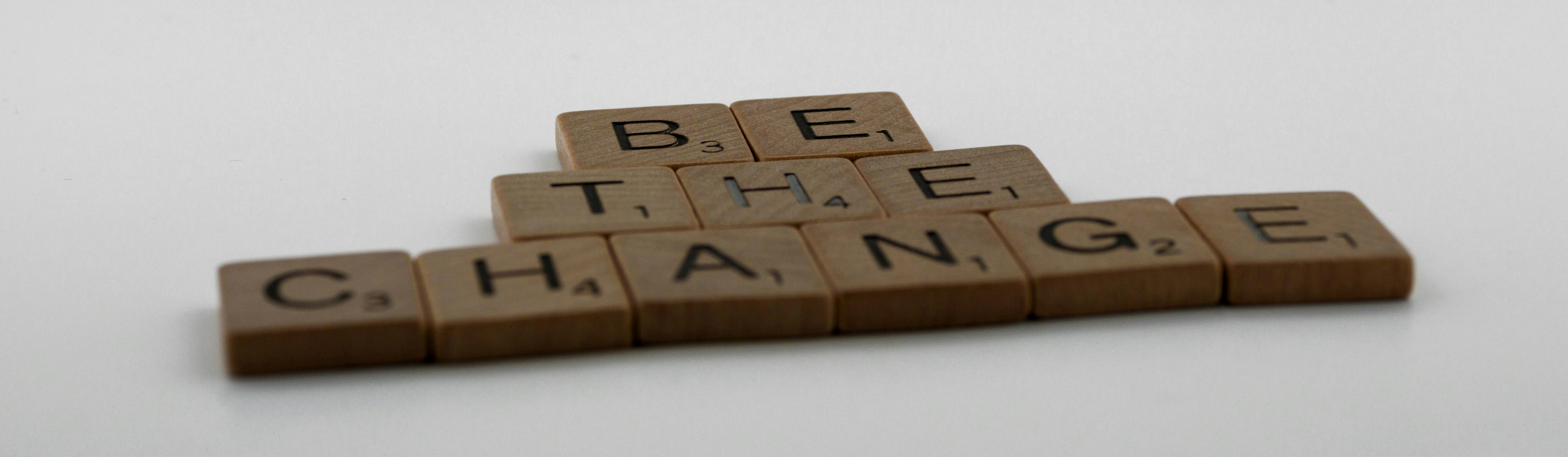}
\caption{Be the change, Brett Jordan, Unsplash, 2024.}
\label{fig:teaser}
\end{teaserfigure}

\received{29 March 2024}

\maketitle

\section{Introduction}
The role of Software Engineers is changing. With the digital transformation and the rise of AI, software engineers (as everyone else) are affected by short- and far-reaching changes for example by changing daily work practices (using LLMs for coding), but also a rise in complexity when developing new software due to the constantly changing requirements resulting in new features and the emergence of new technologies. Moreover, software engineers are not simply execution wizards, even if they sometimes see themselves that way ~\cite{betz2022software}. They are working in all phases of the IT product and service development process. Starting from the early requirements to the maintenance phase ~\cite{SWEBOK2014}. Moreover, nowadays sustainability issues demand responsible design decisions as IT products and services have short- and far-reaching impacts on societies and environment ~\cite{becker2015requirements}. Thus the role of Software Engineers is changing and continues to do so. This means Software Engineers need to adapt (their role) both quickly and in the long run. However, this is not happening. David Gotterbarn already pointed out the lack of education for moral responsibility of software engineers back in 1995~\cite{gotterbarn1995moral} and became one of the champions in developing the codes of conduct in our field. Almost 30 years later we are still primarily educating future professionals for a commercial market instead of for responsible citizenship --- while that is more urgent than ever because of software ruling our daily lives ~\cite{peterson2023abstracted, storey2020software, becker2023insolvent}. This lack of reflection in computing and computing education is also criticized by Becker in his recent book ``Insolvent''~\cite{becker2023insolvent}. Becker argues that the current (dominant) way of thinking in computing is not sufficient to address the grand challenges we face. He suggest to reorient design in computing to better align with the values of sustainability and justice ~\cite{becker2023insolvent}. Moreover, in ~\cite{betz2022software} Betz et. al. show in a small interview study that the current perception of Software Engineers regarding their role is still very classical concentrating on executive tasks such as coding. To summarize, there is a need for research with regard to current challenges in terms of complexity, sustainability as well as responsibility and how the role of the software engineer changes along with the ever-increasing impact of their profession on everyday life. However, to the best of our knowledge, there is not much work to be found on the role of software engineers and their perception of it. This gap is also visible in the small number of ethnographic studies in software engineering~\cite{sharp2016role, zhang2019ethnographic}, a research method that serves well to investigate human aspects. In the words of Gotterbarn: ``Professionalism maintains a proactive posture.'' This paper calls for and proposes a very active stance in educating software engineers for their multi-fold role and turning them into stewards of their profession beyond commercial practice. Our \textbf{contribution} entails (1) a \textbf{vision} for a line of research investigating the understanding of the (ever changing) role of software engineers via ethnographic studies, (2) an initial \textbf{framework} of how to include such understanding in education within a general Software Engineering curriculum to infuse it as opposed to singling it out into a (often optional) elective course, and (3), most importantly, a \textbf{continued discourse} to have an informed conversation with self-reflection and iteratively improved understanding.

\section{Background and Related Work}

This section details suitable research methods as well as background and related work on roles and education, responsibility and ethics, and the role of AI.

\subsection{Suitable Research Methods}
The purpose of ethnographic studies as a technique is to study the community of software engineers and improve the way in which they work. The field of SE has not been very strongly engaged in Ethnography when it comes to roles of software engineers, see ``Role of ethnography in empirical SE''~\cite{sharp2016role}. There are four main features of ethnographic research: (1) the informants' point of view, (2) the ordinary detail of life as it happens, (3) the interpretation and analysis, and (4) a comprehensive and detailed set of data (rich picture).
In addition, there are five dimensions of ethnographic studies: (1) observation, (2) field study duration, (3) space and location, (4) theoretical underpinnings, and (5) ethnographer's intent.

The paper at hand proposes to and advocates for using ethnography as listed in~\cite{sharp2016role} to (1) To strengthen investigations into the social and human aspects of software engineering, (2) To inform the design of software engineering tools, (3) To improve process development and role definition, and (4) To inform research programmes --- in this case, for the purpose of shedding light on the new responsibilities.

\subsection{Roles \& Education}
The proof of concept document of the Guide to the Software Engineering Body of Knowledge (SWEBOK\footnote{\url{https://www.ieee.org/about/ieee-india/ieee-computer-society-india/swebok.html}}) was made available in 1998 - now more than 25 years ago. While SWEBOK does not explicitly define the role of a software engineer, it defines tasks and activities and it does state ``A software engineer displays professionalism notably through adherence to codes of ethics and professional conduct''~\cite[Chp.~11,~p.~2]{SWEBOK2014}. Garousi et al.~\cite{garousi2019understanding} identify knowledge gaps in software maintenance, software configuration management, and testing. 

To understand the role of a software engineer, Li et al.~\cite{li2015makes} performed 59 semi-structured interviews, spanning 13 Microsoft divisions, including several interviews with architect-level engineers with over 25 years of experience. The contribution of their effort is a thorough, specific, and contextual understanding of software engineering expertise, as viewed by expert software engineers. Mary Shaw's roadmap~\cite{shaw2000software} proposed challenges for educators of software developers to help identify aspirations for software engineering education. As software engineering educators, we have been working on a clarification of the roles involved in software development and Shaw points out that challenges are: appropriate credentialing for those roles; an improved treatment of engineering issues; faster response of educational content to changes in technology and fundamental understanding; and better use of information technology in our own education and training. 
In this line of research, we are contributing to a further, and updated, clarification of the role of a software engineer as well as the continued change that comes with this profession and its tasks.
Peters et al.~\cite{peters2023sustainability} performed a systematic review of the literature on sustainability in computing education. From a set of 572 publications extracted from six large digital libraries plus snowballing, the authors distilled and analyzed 89 relevant primary studies. They investigate (i) conceptions of sustainability, computing, and education; (ii) implementations of sustainability in computing education; and (iii) research on sustainability in computing education. The result is a framework capturing learning objectives and outcomes as well as pedagogical methods for sustainability in computing education.
We rely on these results for some of the arguments presented in the paper at hand.

In their paper Betz et. al~\cite{betz2022software} investigate how Software Engineers perceive their role, how they attribute themselves and their awareness with regard to sustainability and they found out that Software Engineers seem to see themselves mainly as executing force.







\subsection{Responsibility and Ethics}
There is a lot of different keywords (and thus work) to be found in this area (responsibility, ethics and software engineering): accountability, ethics-aware and ethical-driven, responsibility, codes and ethics. We look at selected exemplary work in each of it.

David Gotterbarn was the first to take a strong stance for the moral responsibility of software engineers~\cite{gotterbarn1995moral}, noting ``a surprising lacuna in software engineering education, one of whose primary foci is the preparation of software professionals for the commercial market''. However, it seems there is still a lack of (operational) research in this area. In their recent mapping study (2023) Marebane et al.~\cite{marebane2023mapping} investigate the studies on ethical responsibility of software engineers. They only found 14 relevant papers, most of them being philosophical studies and no validation or experience papers. 

With regard to accountability Schneider and Betz~\cite{schneider2022transformation2} wrote that software engineering needs to become accountable for sustainability. In order to do so they combine a social scientific sustainability transformation model with the analysis of accountability for sustainability in current software engineering practices.  They argue that for software engineers to become accountable for sustainability a mind shift needs to be established~\cite{schneider2022transformation2}. This mind shift naturally effects also the roles of software engineers. Aydemir and Dalpiaz~\cite{aydemir2018roadmap} present in their paper a roadmap for an analytical framework for ethics-aware Software Engineering. In ethics-aware Software Engineering the idea is that the ethical values of stakeholders (including software engineers and users) are taken into account during the software requirements engineering and development phase. The framework identifies the subject (what), the threatened object (who) and the relevant value (what). Although, this framework seems quite helpful, it does not touch upon the issue of changing ones role. In their philosophical paper, Genova et al.~\cite{genova2006ethical} argue that as a practical guide for ethical responsible software engineering one should stick to a moderate deontologism taking into account rules and consequences when assessing the ethical impact of software engineering practices. They argue that this is due the need of assessing possible short and long term impacts of design decisions (consequences) while taking into account “absolute human values” (rules). However, they do not provide methods and tools for the assessment. Lurie and Mark~\cite{lurie2016professional} propose a practical ethical framework for software engineers connecting software engineers’ ethical responsibilities directly to practical activities to overcome the dichotomy between professional skills and ethical skills. They call the approach Ethical-Driven Software Development (EDSD), as an approach to software development. The goal of the framework is not to affect the outcome of the software engineering process but to raise software engineers awareness regarding the “ethical ramifications of certain engineering decisions within the process.” (~\cite{lurie2016professional}, page 1). Awareness is definitive a first step, but as following steps we need assessment of impacts on ethical issues.  In their paper from 2022 Mitchell et all. argue based on an interview study that there exists a ``lack of tool and process support for systematic ethical deliberation at most stages of the software lifecycle.''~\cite[p.~1]{mitchell2022incorporating}. To summarize, as Karim et al.~\cite{karim2017ethical} state despite a sufficient large number of research on Software Engineering Codes of ethics operationalising and implementing those codes in the daily work practices is quite difficult and not yet solved. We argue in order to do so we need to change the software engineering roles and their responsibilities, tasks and activities.


\subsection{The Role of AI}

\begin{figure*}[htb!]
    \centering 
    \includegraphics[width=\textwidth]{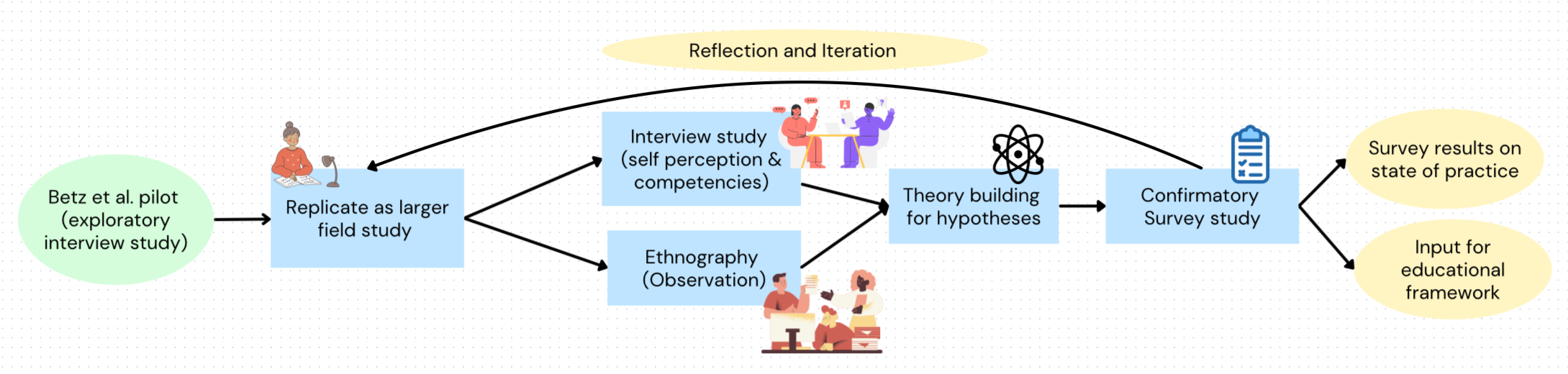}
    \caption{An overview of the planned line of research}
    \label{fig:researchline}
\end{figure*}

The first paper to address AI and Software Engineering is already from 1988, and it looks at attempts to apply AI techniques to software engineering problems~\cite{barstow1988artificial}. This technology has increased the scope of our research tremendously, and with that also increased the responsibility of software engineers. Currently, there is a large amount of work exploring the uses of AI and less work taking a critical perspective on responsibility around AI within SE. Feldt et al.~\cite{feldt2018ways} provide an AI in SE Application Levels (AI-SEAL) taxonomy. The taxonomy has three facets allowing its users to classify, the point of application (process, product and runtime), the type of AI technology (based, initially, on the five tribes proposed by Domingos~\cite{domingos2015master} and the level of automation of the applied technology (inspired by Sheridan-Verplanck’s 10 levels of automation~\cite{sheridan1980computer}). In their very recent paper Pant et al.~\cite{pant2024ethics} investigate ethics in AI from an IT practitioners point of view. They developed a taxonomy to help AI practitioners in identifying and understanding AI ethics. Moreover, they presented recommendations for IT industry and future research. For example, they advocate examining the different roles and their interactions in the development and use of AI systems.

In light of most current research and development, we naturally want to include AI in the discussion, but it is not the main focus of the work in the paper at hand.

\section{Future Research Challenges}

We as software engineering researchers need first to understand what the role currently is in the understanding of the IT practitioners (bottom-up, ethnography) and at the same time, secondly address it top-down via education. Thirdly, we wish to establish it as part of the culture of software engineering to reflect upon the role of its professionals. Along with this reflection and the awareness of the influence of software engineering on society, the role of Software Engineering changes iteratively and in correspondence to the overall technological development. In order to do so, we get to work bottom-up and top-down in parallel.

\subsection{Vision for a Line of Research}
We present a vision for a line of research investigating the understanding of the (ever changing) role of software
engineers via ethnographic studies. 

According to Stol's ABC~\cite{stol2018abc}, the right place to study the current role of software engineers for the purpose of studying actors and their behaviour in natural contexts is the `jungle': ``Natural setting that exists before the researcher enters it. Minimal intrusion of the setting so as not to disturb realism, only to facilitate data collection.''~\cite[p.~13]{stol2018abc}. Sharp's review of ethnographic studies~\cite{sharp2016role} reveals that there is a relatively small set of them within software engineering, most likely due to the fact that pure ethnography is assigned little value in software engineering before also offering the evaluation of an intervention (field experiment).
Consequently, Step 1 of the research vision is an ethnographic study of how software engineers currently deal with ethical issues. Betz et al.~\cite{betz2022software} find that software engineers still see themselves mainly as executive coders that implement as they are being asked to without questioning the specifications too much. Since this study was based on a small sample, the first step in the vision at hand is to replicate the study with a larger number of interviewees. This larger study will give us the hypotheses upon which we can base field experiments. 

Based on an understanding of this current view of their (perceived lack of) responsibility, we envision to roll out a larger field experiment in order to investigate how software engineers deal with cases presented that have ethical concerns, see Fig.~\ref{fig:researchline}. The research line includes an extended interview study as well as an observational study that are both used for input for theory building. The theory will be evaluated in a confirmatory survey study and these results will, in turn, inform about the state of practice and serve as input for the educational framework for needed competences. Moreover, iteratively applied the planned line of research serves as an input for the continued discourse detailed in Sec.~\ref{sec:framework}.

\subsection{Initial framework for new activities, tasks and responsibilities of software engineers} 
To identify the needed new activities, tasks and responsibilities defining the new role of software engineers, we present here three abstraction levels: Competencies, method (refining competencies) and technologies (refining methods). They help us think through this improved understanding of roles (see Table~\ref{tab:mapping}). 

\begin{table}[htb]
    \centering \footnotesize
    \begin{tabular}{p{2.5cm}|p{2.5cm}|p{2.5cm}}\hline 
        \textbf{Competency} & \textbf{Method} (refines competency) & \textbf{Technology} (refines method) \\\hline \hline 
       Long-term thinking~\cite{becker2023insolvent}  & Systems thinking and critical thinking~\cite{becker2023insolvent} & Soft systems methodology ~\cite{checkland2020soft}\\\hline
       Complexity theory~\cite{mcdaniel2001complexity} & Consideration of and working in different and between different abstraction levels & Agent based modelling ~\cite{manson2012agent} \\\hline
       Environmental externalities and impacts~\cite{harris2021theory} & Life cycle analysis~\cite{ciambrone2018environmental}, Sustainability Assessments~\cite{bond2012sustainability} & AI background knowledge\\\hline
       Diversity Equity Inclusion & Ethnography~\cite{sharp2016role} & Data science understanding, bridging the digital divide\\\hline
       Ethical concerns~\cite{genova2006ethical} & Ethics as a skill~\cite{jarvinen2017ethics} & Modelling Ethics for SEs~\cite{SabahModellingEthics4SE}\\\hline
    \end{tabular}
    \caption{Needed competencies, methods and technologies for a changing role}
    \label{tab:mapping}
\end{table}

For example in order to consider long-term thinking during the IT product and service development, one should apply systems thinking and critical thinking using for example the soft systems methodology. In this case the software engineering's role would be extended with regard to responsibilities by applying long-term thinking to raise awareness regarding sustainability issues as well as activities and tasks supporting it (e.g. a soft systems modelling exercise). 

This is an initial framework, where the different wording and methods and tools are only a proof of example and can be discussed. We plan to further investigate and extend this based on a literature study and the ethnographic study described above. We started with the provided competences as Long-term thinking is crucial for understanding and assessing the current Grand Challenges faced by humanity. We added Complexity Theory as Software Systems become more and more complex also given the current AI trend. Environmental externalities have been included as the climate change is affecting all and IT Products and Services is one way to minimize negative impacts of it. Finally, we included Diversity, equity and inclusion to support the integration and leverage diverse perspectives that promote innovation and collaboration. and ethical concerns to address responsible  citizenship.

\subsection{Continued Discourse on Self-Reflection} \label{sec:framework}

As one aspect of lifelong learning as software engineers~\cite{chernenko2023lifelong}, the authors of this paper believe we need an iteratively improved understanding of our role in this profession and the responsibility that comes with it. Explained in stages that use the terminology of the revised Bloom's taxonomy~\cite{forehand2005bloom}, we propose to approach a continued discourse and self reflection of one's work practice as described in Table~\ref{tab:discourse}. We see the six levels of the revised Bloom's taxonomy and an explanation of their basic understanding in the left column, then an example activity in the middle column, and more details about how this could be carried out within an Software Engineering com in the right column.

\begin{table}[htb]
    \centering \footnotesize
    \begin{tabular}{p{2.5cm}|p{2.5cm}|p{2.5cm}}\hline \hline
        \textbf{Bloom's} & \textbf{Activity} & \textbf{Details} \\\hline
        Remember: recall facts and basic concepts & Training for sustainability awareness & Seminar during onboarding plus yearly update \\\hline
        Understand: explain ideas or concepts & Exemplary case study discussions & Solidify concepts in quarterly training \\\hline
        Apply: use information in new situations & Weekly reviews with personal reflections & Ensure cognisance of ongoing themes in self reflection e.g. report in daily stand up meetings if ethical issues occur\\\hline
        Analyze: draw connections amongst ideas & Monthly reviews of personal reflections e.g. as part of the sprint retrospective & Aggregate higher level reflection to identify potential patterns \\\hline
        Evaluate: justify a stand or decision & Reviews for ethical issues in company using focus group meetings & Take notes to collect a database of issues and how they are investigated, tracked and decided e.g. include ethical issues in user stories \\\hline
        Create: produce new or original work & Guidelines for individual company & Develop a set of tailored guidelines for one's company \\\hline 
    \end{tabular}
    \caption{Continued discourse and self reflection of one's work practice}
    \label{tab:discourse}
\end{table}

\section{Conclusion}

We presented a threefold contribution: (1) a Roadmap with ethnographical studies to investigate state of practice, (2) a Framework for the new understanding of the role of a software engineer, and (3) a Continued Discourse on Self-Reflection. For the near future, we plan to sketch out the details of the first ethnographic study as well as the interview study. Closing this paper we want to raise one last point: Is this really new? We think yes, to a certain point. We acknowledge, that it is a repetition of a call-to-action since we haven't taken sufficient action. However, the path has changed and is new - we need to walk it via a change in the understanding of our role as software engineers.








\bibliographystyle{ACM-Reference-Format}
\bibliography{bib}

\end{document}